%% file: Main-ProblemSpace.tex
\documentclass[fleqn,10pt]{SelfArx} 

\usepackage[english]{babel} 

\usepackage{multicol}
\usepackage{adjustbox}
\usepackage[final]{changes}

\usepackage[numbers]{natbib} 

\usepackage[resetlabels]{multibib}  
\newcites{Side}{\raggedright{\small{References Sidebar Task Abstractions}}}

\definecolor{color1}{RGB}{0,0,90} 
\definecolor{color2}{RGB}{0,20,20} 

\usepackage{hyperref} 

\hypersetup{
	hidelinks,
	colorlinks,
	breaklinks=true,
	urlcolor=color2,
	citecolor=color1,
	linkcolor=color1,
	bookmarksopen=false,
	pdftitle={Title},
	pdfauthor={Author},
}

\JournalInfo{Author's submitted version} 
\Archive{} 

\PaperTitle{A Problem Space for Designing Visualizations} 

\Authors{Michael Gleicher\textsuperscript{1}, Maria Riveiro\textsuperscript{2}, 
 Tatiana von Landesberger\textsuperscript{3}, Oliver Deussen\textsuperscript{4}, Remco Chang\textsuperscript{5}, and  Christina Gillman\textsuperscript{6}} 
\affiliation{\textsuperscript{1}\textit{University of Wisconsin - Madison, Madison, WI, 53706, USA}} 
\affiliation{\textsuperscript{2}\textit{J\"onk\"oping University, J\"onk\"oping, Sweden}} 
\affiliation{\textsuperscript{3}\textit{Universit\"at zu K\"oln, K\"oln, Germany}} 
\affiliation{\textsuperscript{4}\textit{University of Konstanz, Konstaz, Germany}} 
\affiliation{\textsuperscript{5}\textit{Tufts University, Medford, MA, 02155, USA}} 
\affiliation{\textsuperscript{6}\textit{{L}eipzig University, Leipzig, Germany}} 

\Abstract{
\\Visualization researchers and visualization professionals seek appropriate abstractions of visualization requirements that permit considering visualization solutions independently from specific problems.
Abstractions can help us design, analyze, organize, and evaluate the things we create. The literature has many task structures (taxonomies, typologies, etc.), design spaces, and related ``frameworks'' that provide abstractions of the problems a visualization is meant to address. 
In this viewpoint, we introduce a different one, a \emph{problem space} that complements existing frameworks by focusing on the needs that a visualization is meant to solve. We believe it provides a valuable conceptual tool for designing and discussing visualizations.
\thanks{Author's submitted version. An article with the same content was approved for publication in the Visualization Viewpoints Department of IEEE Computer Graphics and Applications magazine. This preprint will be updated upon publication of that article.}
}

\newcommand{\keywordname}{Keywords} 

\begin{document}

\maketitle 
\thispagestyle{empty} 

\section*{Introduction} 
A \textit{problem space} is a conceptual structure (i.e., a type of abstraction) that helps organize thinking  about the problem to be addressed, sometimes referred to as needs (e.g., \citet{olsenChapterProblemSpace2015}).

The design and product development literature provides various schemes that encourage consideration of the problem to be addressed independently from the solution that will address it; this separation is the core tenet of a problem space.
Proponents argue that keeping these two separate has advantages that include focusing on solving correct problems, allowing more creative searches for solutions free of preconceived ideas, and better articulation of needs to enable communication and checking. In this viewpoint, we bring the concept of a problem space to the design of visualizations (the creation of anything from specific charts to analysis systems) by providing an abstract framework for considering visualization problems.

We articulate an abstract problem space, i.e., a form of abstraction, 
to provide a conceptual tool for organizing thinking about the needs for a ``visualization,'' independent of its ultimate design. Arguably, the premise that the solution will be a visualization is conflating problem and solution; we take a broad view of what a visualization may be, as it could be that the right solution to a problem may not require any visualization. It might be better to call this a problem space of data analysis and communication problems. \added{We believe our work complements existing abstractions for visualization design, and provide context in the sidebar.}

\begin{table} 
    \centering        
    \caption{The six axes of the problem space provide aspects to consider in design.}
    \begin{tabular}{l p{6cm}}
        \toprule
        \textsc{\textbf{Who}}  &   Who has the problem? Who will use the visualization? \\
        \textsc{\textbf{Why}}  &   Why do they need a visualization? What is the objective? \\
        \textsc{\textbf{What}} &   What is the data?\\
        \textsc{\textbf{When}} &   When in the analysis will they use the visualization?\\
         \textsc{\textbf{Where}} &  Where will they use the visualization? What is the context?\\
         \textsc{\textbf{How}} &     How do they expect to be helped?\\ \hline
        \bottomrule
    \end{tabular}

    \label{tab:aspects}
\end{table}

Our abstract problem space provides a set of different aspects, listed in Table~\ref{tab:aspects},  that should be considered in devising a visualization solution.
Briefly, a visualization is intended to help someone (the \emph{who}) do something (the \emph{why}) with the data (the \emph{what}) in some phase of the analysis process (the \emph{when}) and in some context (the \emph{where}) using some method (the \emph{how}).

\added{Note that \emph{how} refers to user's expectations of a solution, which is an aspect to be considered in design, not the ultimate solution, which should not be part of the problem space.}
This list of questions serves as the dimensions of our problem space. 
It provides a checklist to remind us of factors that we need to consider when designing a visualization. It provides a vocabulary to articulate the problems in a consistent way, allowing us to communicate problem needs and to later assess how a design considers them. It provides a way to organize aspects so that similar problems, and their solutions, can be identified.

We consider the problem space using the metaphor of mathematical spaces. Each aspect forms an ``axis'' - a generally independent direction in the space, such that any specific ``problem'' (a scenario that we might want to create a visualization for) is a point along that axis, and therefore a point in the high-dimensional problem space. Each axis represents a range of responses; they may be a continuous range (truly an axis), 
they may be multi-faceted with multiple dimensions that provide a space of responses, or they may be simplified to a set of discrete categories.

We have developed the problem space based on our experience creating visualization solutions. It provides a checklist of aspects to consider in visualization design, and serves as a pedagogical tool in teaching designers about factors they may need to consider.

\section*{Background}
Our problem space is motivated by 
the work of \citet{Schulz2013}
who applied the idea of spaces to the abstraction of task in visualization.
However, like \citet{rindTaskCubeThreedimensional2016}, we observe that across 
the rich literature on ``task'' in visualization
there is substantial variance in what the term
may mean (see sidebar). 
Across the literature, ``task'' may be used as a term for anything ranging from a specific thing that a viewer is trying to do to a more general context of what will be done and why. We believe that all of these are useful concepts, and try to use the problem space to capture the broadest sense, and use particular aspects of the space to capture the more focused notions of task. 
Our problem space seeks to be broader (e.g., including many notions of task), but also differently limiting (i.e., focusing on problems at the exclusion of their solution).

\citet{Schulz2013} applied the ``5 Ws And How framework,'' common in journalism and general information seeking \citep{FiveWs} to create a ``Design Space of Visualization Tasks.'' 
In our experience applying this work to our own design and teaching, we find that their focus on providing a formalism for task was less appropriate for providing general guidance on the range of aspects one must consider when designing a visualization.
We also apply the 5Ws and H as our framework, but with a broad focus on the problem without trying to presuppose the solutions. We include aspects we feel allow us to describe problems in ways that are independent of the ultimate solutions.

The 5Ws and How framework provides information seekers six questions to ask (or answer, in the case of journalistic writing): Who, What, When, Where, Why and How \citep{FiveWs}. While the 5Ws and How are commonly referred to in journalism to help writers organize key information, the idea has historical roots that date back (at least) to Aristotle's rhetoric. Intriguingly Aristotle argued that ignorance of any of these elements can lead to faulty reasoning. We believe that Aristotle's advice applies to visualization designers: each question is something designers should consider. An insight of the work by \citet{Schulz2013} was to define the questions in a Visualization specific manner, a task we set out to do here to create our problem space.

\section*{The Axes of the Problem Space}

The major axes (aspects) of our problem space consist of the Five Ws and How, introduced in Table~\ref{tab:aspects}
and detailed in this section. Admittedly, we are forcing these to map to the 5Ws and How, which may feel a little contrived. However, we believe that the history and mnemonic power of the framework make it worth sticking to this set of words. 
The six aspects provide a set of considerations for design.

\begin{figure*}[htp]
    \centering
    \includegraphics[width=\linewidth]{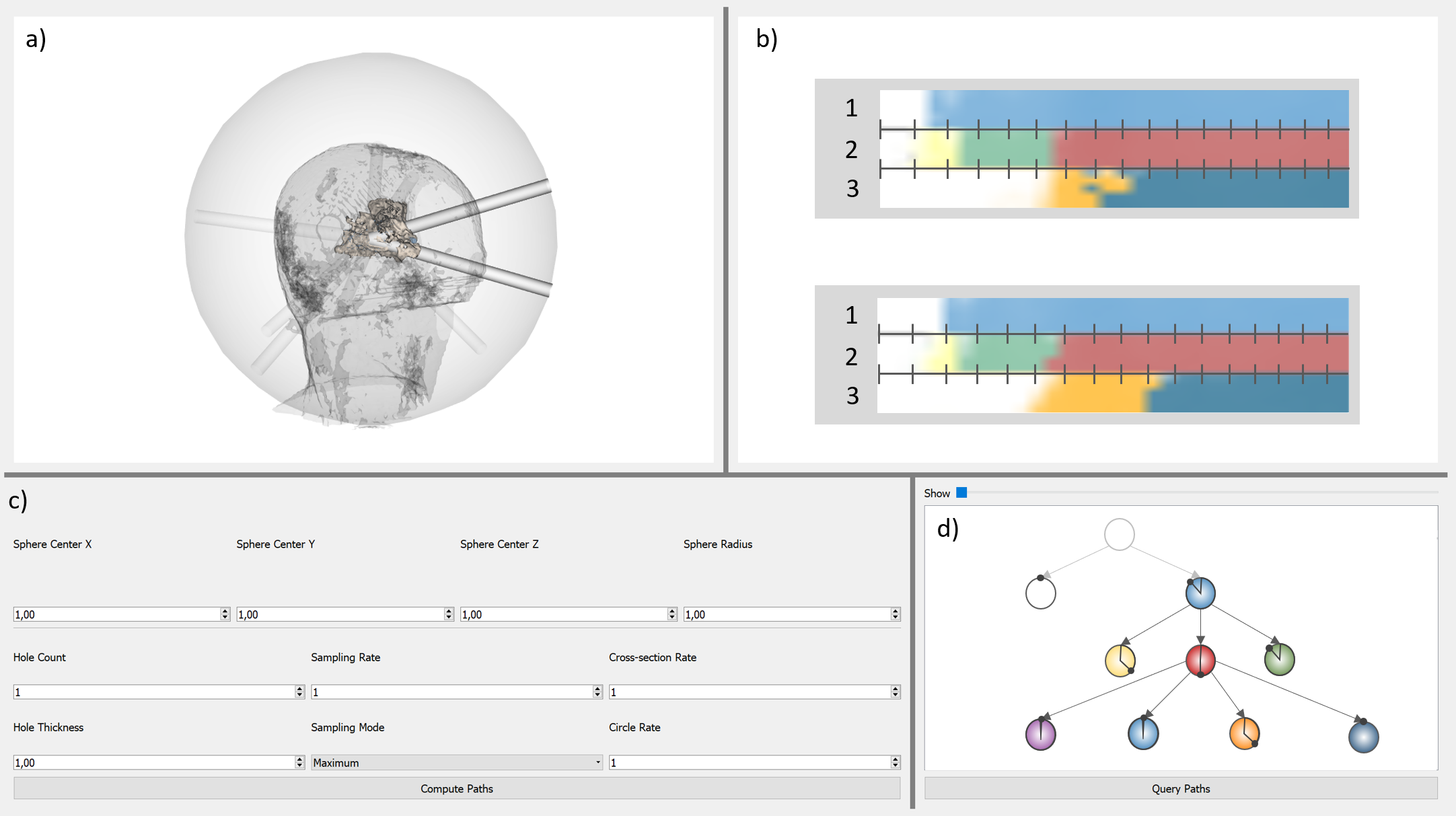}
    \caption{Running example from \citet{Gillmann2018} to demonstrate the use of the presented problem space for designing visualizations. The tool presents a visualization to assist keyhole surgery planning. It is composed of a volume view (a), a surgery tunnel view (b), a control panel (c) and a control graph to enhance specific areas in the human brain (d).}
    \label{fig:example1}
\end{figure*}

To help make our abstractions concrete, show the usability, usefulness and the generalizability of our problem space definition, we provide example use cases.
We selected the use cases across data types and application domains. 
The first use case, used as a running example in this section, considers a keyhole surgery planning with the goal to remove a tumor from a patient's brain \citep{Gillmann2018}, which can be found in Figure \ref{fig:example1}. Two other use cases are provided in a later section.
We acknowledge that the examples apply the framework \emph{post-hoc} to recent projects that were completed prior to the articulation of the framework. Our experience with these, and other projects, were an important basis for developing the framework.
\added{Details of the use cases can be found in their original publications, we only describe aspects related to problem dimensions.}

\noindent\subsection*{Who has the problem?} Generally, our focus is on the intended audience of the visualization; the people who have some need to see something in the data. Many aspects of the audience might serve to influence the solution. For example, the level of analytic sophistication of the expected viewer, their level of motivation, their spatial reasoning abilities (e.g., \citet{ottleyImprovingBayesianReasoning2016}), or their background knowledge. Considerations such as visual acuity, or accessibility (e.g., supporting viewers with color vision deficiency) are also part of the \emph{Who} aspect.

Design may also consider that problems have multiple stakeholders. While the focus on visualizations is often the ultimate viewers/users, other stakeholders involved in the problem may need to be considered in the choice of a solution. There are always the developers and maintainers to consider, and in some cases even the ``users'' might involve a complex set of analysts, decision makers, writers, and readers.

In the running example, the \textit{who} is the clinician who aims to process a keyhole surgery. For communication reasons, the who might also include the patient that will be treated by the keyhole surgery. In terms of design, this led to a multi-view system that covers different aspects of the keyhole surgery in terms of different types of tissue that are interfered with by the procedure, as shown in Figure~\ref{fig:example1}. The views were chosen to be familiar to the clinicians, who are medical experts.

\noindent\subsection*{What is the data being worked with?} We assume that in a ``data analysis or communication problem'' there is data to be visualized (although obtaining the appropriate data may be part of the problem). The data impact the solutions in many ways and at many levels.

Many different aspects of the data can influence the solution. The low level details (i.e., the ``form'' or type of the data) have obvious implications for the choice of solutions. However, there is an increasing acknowledgment that higher level factors in the data, such as its semantics, distribution, value ranges, important relationships, etc. need to be considered in design \citep{walny2019data}. Issues such as quantity and time-dependence of data can influence solutions: e.g., data that is very large, or changing rapidly requires design consideration.

The input or provided form of the data may differ from its internal representation, or even how the user may think of it. Data may be transformed as part of the analysis process, or thought of in different ways. Any of these considerations may impact design. 

In the running example, the \textit{what} is a Computed Tomography (CT) scan of the patient's head. In addition, the tumor, that needs to be removed is known by the clinician and indicated in the CT scan. Further, potential surgery paths that can be used to process the procedure are part of the available data. In terms of design, this led to a dedicated view for different surgery tunnels that are classified to be suitable, as shown in Figure \ref{fig:example1} b).

\noindent\subsection*{Why is the user working with the data?} Here we consider the goal of the visualization. Typically, this is \emph{what\footnote{We use \emph{what} for the object of exploration, so this question is ``\emph{Why?}} is the viewer trying to do with the data?''} but the nature of this goal may change as multiple stakeholders are involved (e.g., it may be what the author intends for the audience to achieve with the visualization).

The \emph{Why} most closely aligns with the common notion of task, with the caveat that many discussions of task do not separate it from the other aspects we describe (e.g., \citet{Schulz2013} who include all 5Ws and H as part of task). \citet{rindTaskCubeThreedimensional2016} distinguish perspectives on task as ``why'' and ``how'' with the terms \emph{objectives} and \emph{actions}. Their use of Roth's concept of objectives \citep{rothEmpiricallyDerivedTaxonomyInteraction2013} aligns with our goal of separating the problem from the solution.

Many challenges of discussing tasks abstractly, such as the need to consider multiple levels of abstraction \citep{Brehmer2013}, must be considered.
The question \emph{Why?} can provide a useful tool in traversing the levels of abstractions and separating problem and solution: if a description seems too specific, the question ``Why?'' often leads to a useful generalization.
Conversely, the question ``How?'' can be used to refine an overly general goal, for example, to break it into subgoals or give a more concrete target. However, this question might also invite conflating the solution with the problem goal.
\added{We can apply many of the existing works on task descriptions (see sidebar) to articulate tasks. However, within the framework, we look to focus on the problem, without presupposing a solution, and to separate the \emph{Why} as much as possible from the other aspects.} \added{We also acknowledge that there may be multiple tasks involved in a particular scenario.}

In the running example, the \textit{why} can be phrased as the task to determine the safest and most suitable surgery path for the planned keyhole surgery. In order to remove the tumor of the patient, healthy tissue needs to be interfered with to reach the tumor. Here, the \textit{why} expresses the task of determining the path that destroys as little healthy tissue as possible while being able to reach and remove the tumor. Therefore, the presented visualization allows 
the clinician to determine which areas \added{should} be avoided while choosing their path.
This can be done by the user using the control panel, the control graph as shown in Figure \ref{fig:example1}.

\added{The example highlights the importance of choosing an appropriate level of abstraction for a task in design. In principle, the surgeon most likely has ``higher level'' goals (such as promoting the health of their patient), and lower level goals (such as making good use of their time or interpreting the available imagery). While many different tasks may be identified, a designer should focus on the ones that are likely to inform the design and evaluation.}

\noindent\subsection*{When in the process does the problem occur?} In this aspect, we acknowledge that ``data work'' is a process that may include gathering information, defining questions, performing analysis, drawing conclusions (or acting on the analysis), communicating or justifying rationales, or other steps. \replaced{The process varies, but is often}{This process is} iterative and not always sequential. Understanding where along this process the visualization may be used can inform design: curating data or performing analysis, or making decisions, may benefit from different choices than communicating an identified result or justifying a decision.

Different phases of data usage may suggest different choices for visualization design. For example, a visualization used for \added{the} initial examination of the data may prefer more overview and exploration than one designed to communicate specific findings later in the process.

In the running example, the question \textit{when} can be answered with: before the keyhole surgery. As described, the when can change. In our example, this could mean that there could be a different visualization assisting the \replaced{surgeon}{surgent} during the surgery to ensure that the planned surgery path is followed or even another visualization after the surgery to examine if the procedure was successful. In the presented example all visual components are designed to target the timeframe before the procedure and allow the clinician to explore different options.


\noindent\subsection*{Where does the problem occur?} For this axis we consider the \textit{context} of the problem. \emph{Where} can include the literal place (e.g., the user is in the lab vs. on a bus), 
\added{the user's activity (e.g., is there attention focused on the visualization?),} 
\added{the social environment (e.g., will the user be sharing the visualization?),}
the display they will be using (e.g., a large monitor vs. a cell phone vs. a printed page), or the computing environment (how much computation is available at viewing time). The context should also include where the visualization is used, e.g., is it placed within a newspaper article, a scientific paper, or web page, or is it a stand-alone system that viewers access explicitly? Context might also include the scenario: is the stakeholder making a business decision, evaluating scientific data, etc.? Aspects such as the amount of time available, pressure (high stakes), and the need to justify choices later influence user needs and therefore the ultimate design that addresses them.

In the running example, the \emph{where} is the computer included in the office of a surgeon in a dedicated planning session (i.e., before the surgery, but after imaging and image analysis).
\replaced{The scenario has the user spending focused time with the tool as part of an explicit planning and analysis activity. The clinical context also raises data handling and privacy issues. The available hardware must also be considered as clinics often have older workstations.}
{This implies that the user will spend focused time with the tool, but also requires considering that the equipment may not be up to date.
The clinical context also raises data handling and privacy issues. The presented approach is designed in such a way that older workstations are able to run it.}

\noindent\subsection*{How will the problem be addressed?}
In general, the goal of the problem space is to separate the ``how'' from the other aspects that specify the problem to be solved.
However, it can also be valuable to acknowledge that sometimes these two are not completely separable: the stakeholders and/or designers may have pre-conceived notions of what the solution should be. These should be acknowledged so they can be properly considered as either design constraints, or as biases that might be counteracted. For example, a potential user may demand a familiar visual representation, or an implementer may want to show of an existing tool, whether or not a different solution might better address the problem needs.  In the running example, the design combined views familiar to the physicians (the CT scans) with more novel displays. \added{The novelty in a solution can come from the use of existing components to address a problem in a novel manner.}

\input{sidebar}


\begin{figure*}[htp]
    \centering
    \includegraphics[width=\linewidth]{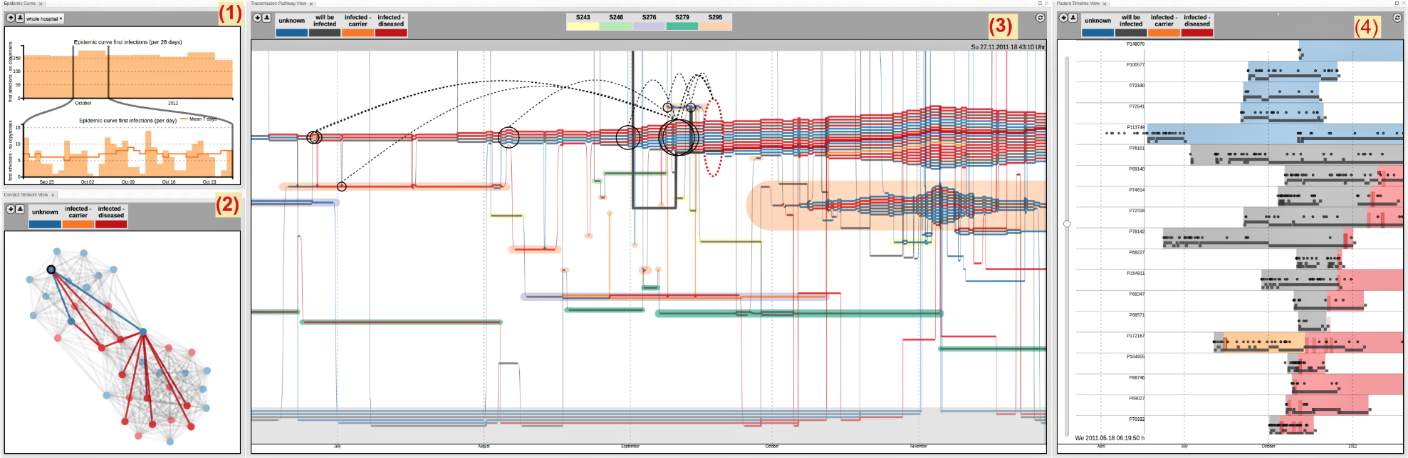}
    \caption{Second example use case from \citet{baumgartl2020search} to demonstrate the use of the presented problem space for designing visualizations. Visual interface for tracing infection control consisting of several views: epidemic curve, transmission pathway view, contact view and patient timeline view. }
    \label{fig:example2}
\end{figure*}

\section*{Additional Examples}
 This section describes two use cases: the tracing of infection pathways in hospitals \citep{baumgartl2020search} with the goal to analyze and monitor infection spreading of multi-resistant pathogens, and the exploration of news articles \citep{yimam2016new,ballweg2016new,muller2017guidance} aiming at finding newsworthy information in large document sets.

\subsection*{Infection Tracing Use Case}

\noindent\subsubsection*{Who:} The exploration of and analysis of infection pathways is conducted by infection control, hygienists, infection control data management, epidemiologists and infection control managers. They are experts in infection control with little knowledge of novel visualizations.
The design tried to support their expertise by integrating familiar views with novel elements.
{They commonly used two types of visualizations: 1) line-charts that show the progress in numbers of infections and 2) manually created patient timelines. Based on this knowledge and expectations, the visual design leveraged and included these two views in the final visualizations. The other views complemented the designs based on common knowledge of node-link diagrams and a storyline that combined several aspects in one view (see Figure~\ref{fig:example2}).}

\noindent\subsubsection*{What:} The data for infection control was complex and large. It consisted of a combination of two datasets: patient locations over time and patient infection status over time. The data comprised hundreds of thousands of patients over several years at a second frequency. This led to a design that focused on one focus patient and her first- and second-level contacts with a multi-scale time resolution. 

\noindent\subsubsection*{Why:} The infection control expert aims at 1) detecting an outbreak, 2) identifying the outbreak pathway (via contacts of patients), 3) determining the location of \added{the} outbreak, 4) quantifying the duration of \added{the} outbreak and 5) identifying potentially infected patients. These goals (intents, tasks) need to be solved using the visualization. Therefore the views are designed so that they enable reaching the analytic goals. Especially, the epidemic curve supports \replaced{task}{Task} 1, the transmission pathway view supports \replaced{tasks}{Task} 2-4, the patient timeline view supports \replaced{tasks}{task} 3 and 4, the contact view supports \replaced{task}{Task} 5 (see Figure~\ref{fig:example2}). 

\noindent\subsubsection*{When:} The data is analyzed after an infection control expert has a hypothesis about a potential outbreak. It is a retrospective analysis to detect, confirm and reconstruct an outbreak in a hospital. It occurs in the analytical phase before reporting to management and health control authorities and before taking any of the infection control measures. 

\noindent\subsubsection*{Where:}  The infection control experts in the office at the hospital with standard hardware (common desktop PC and screen). This needs to be considered and tested in the visualization, however, no responsive visualizations, e.g. for smartphones, were needed.

\noindent\subsubsection*{How will the problem be addressed?} 
\replaced{The infection control experts had experience with existing visualizations, including 1) line-charts that show the progress in numbers of infections and 2) manually created patient timelines. They expected to see the epidemic curve with the exactly defined axes and data and the patient timeline. To leverage their knowledge and expectations, the visual design leveraged and included these two views in the final visualizations. The other views complemented the designs based on common knowledge of node-link diagrams and a storyline that combined several aspects in one view (see Figure~\ref{fig:example2}).}{
The infection control experts expected that we re-use and build upon the visualizations that they used by then. Especially, they expected to see the epidemic curve with the exactly defined axes and data and the patient timeline. They were used to seeing and using these types of visualizations in their analysis. Thus the novel visualization had to include them.}

\subsection*{Journalism -- Exploration of a Document collection -- Use Case}


\subsubsection*{Who:} 
The goal was to support journalists \added{in} exploring collections of documents. These users often work under time pressure, and typically \replaced{were unwilling to invest time in learning new visualizations.}{had low visual literacy.}
Therefore, the visual design could not include complex visualizations but leveraged popular dashboard design with bar charts and a node-link diagram. The need for fast analysis under time pressure was supported by an interactive set of filters and degree-of-interest based data exploration. Moreover, journalists need to work in asynchronous teams with the need to share exploration results with colleagues to continue the exploration. This is supported in the visualization by provenance tracking and sharing functions (see Figure~\ref{fig:example3}).

\subsubsection*{What:} Journalists needed to explore large collections of documents (millions), from which named entities and their relationships were extracted. For example, WikiLeaks PlusD (Cablegate) documents has 1.4 million entities  that are connected by 163 million edges. Showing all the data at once is infeasible so 
linked views with filtering, details on demand and degree-of-interest based data exploration were developed and used in the solution.

\subsubsection*{Why:} 
\begin{figure*}[t]
    \centering
    \includegraphics[width=\linewidth]{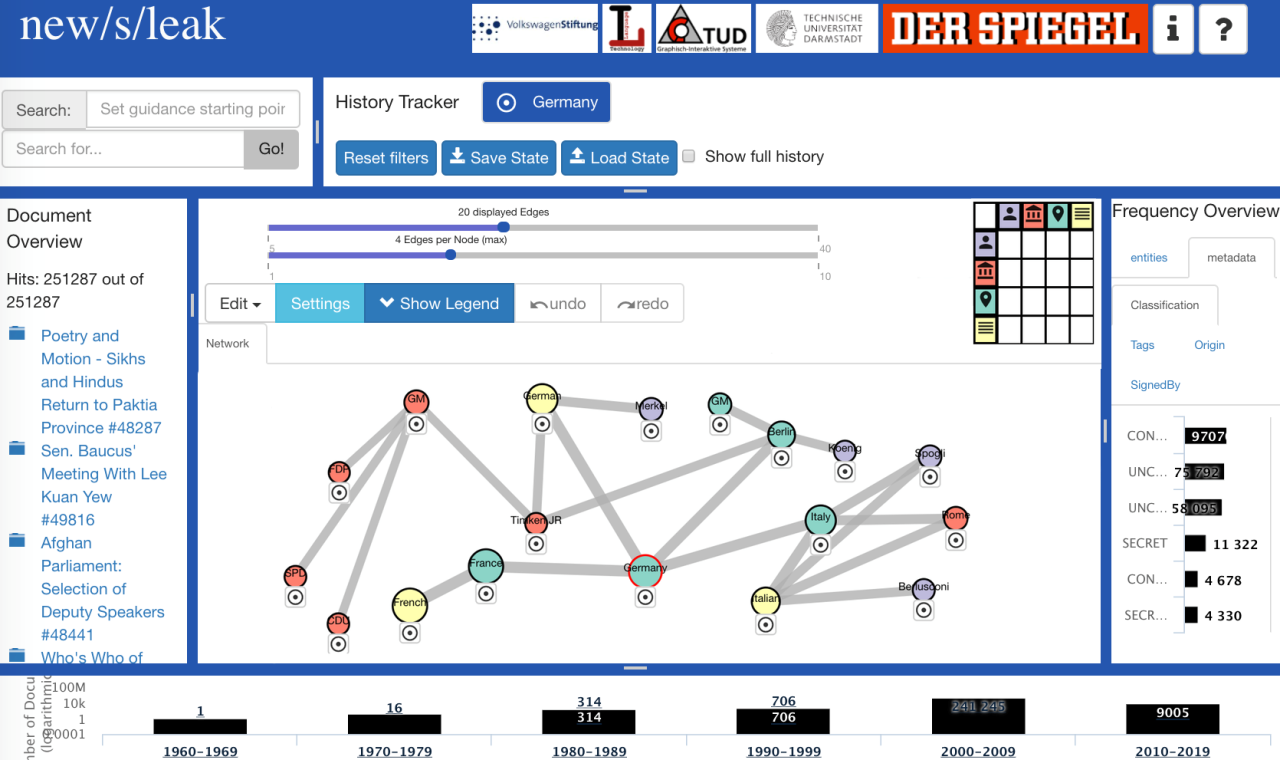}
    \caption{Third running example to demonstrate the use of the presented problem space for designing visualizations. Visual dashboard for investigative journalism based on large document collections. The linked views comprise the distribution and filtering of documents by time and entity and a node-link diagram of entity connections that can be explored using a degree-of-interest function. The exploration process has provenance features in a view on the top \citep{muller2017guidance}}.
    \label{fig:example3}
\end{figure*}

Journalists aim at identifying newsworthy information in large document sets. The visualization allows them to browse and filter document collections by time and named entities, explore links between entities, and read interesting documents in close reading mode. Moreover, it supports data curation and browsing history (see Figure~\ref{fig:example3}). In this way, interesting pieces of information can be found and shared.

\subsubsection*{When:} The data exploration is done during journalistic investigation, in particular, in the early phase of investigation, when interesting facts need to be identified. After their identification using the visualization, these interesting pieces of information need to be checked and confirmed so that they can finally be published. \added{The design focused on the early phases of the journalistic process (e.e., investigation), but required creating results that could be used in later phases (e.g., using the facts to craft the story).}

\subsubsection*{Where:}  The journalists in our use case worked in the office with standard hardware (common desktop PC and screen). Thus responsive visualizations, e.g. for smartphones, were not needed.

\noindent\subsubsection*{How will the problem be addressed?} The journalists posed constraints on the solution, such as expectations to be able to closely read the articles. Summary views and distant reading were not sufficient.

\section*{Is the Problem Space Useful or Good?}

To justify our problem space, we consider a range of desirable properties in such abstractions. These goals are motivated by \citet{kerracher2017constructing} who discuss the evaluation of task abstractions:

\begin{itemize}
\item  \textbf{Relevant} (does each axis make a meaningful distinction): Each aspect was chosen and defined to be relevant to the design and analysis of visualization solutions.
\item  \textbf{Independent} (allows us to discuss problems independently of solutions): We have tried to frame each axis as something that can be discussed in a solution-agnostic manner.
\item \textbf{Thorough} (does it cover all aspects?): we do not assert that our framework is fully thorough, there may be aspects that do not fit neatly into one of our axes. However, we believe that the framework covers an expansive enough range of issues that it is still useful.
\item  \textbf{Complete} (any problem will exist somewhere in our space): Any data problem will involve each of the aspects we describe. In some cases, it might not be specific (e.g., there may be a range of users).
\item  \textbf{Orthogonal} (the aspects are independent of each other): While we have aimed to make the aspects orthogonal, they are ultimately inter-twined or at least correlated. For example, different types of users tend to operate in particular contexts or address different types of tasks. However, we believe that each provides a different lens on what might be the same situation.
\item  \textbf{Precise} (any problem can be uniquely positioned in the space):
We did not provide specific formalisms for how to consider each aspect; this limits the precision of our framework. If precision is important, one can make precise choices for the axes. 
In practice, problems often cover ranges on the axes, such as multiple users or objectives, so precision may not be a valid goal.
\end{itemize} 

\section*{Using the framework}

Our primary premise is that all of these aspects of a problem should be considered as part of a design or evaluation process.

Often, when confronted with a problem we don't know all of the aspects. This uncertainty in the specification of the problem, e.g., not being able to predict who the users will be or where they will be using a visualization, is quite different from uncertainty in the problem itself (e.g., having to deal with noisy data). These unknowns may still be useful: they point to aspects where more clarity might be sought (e.g., better identifying the potential users) or the lack of clarity may influence the design itself (e.g., not knowing the viewers makes for a greater need to support a range of them through inclusive design).

The problem space can serve as a checklist:  have the various aspects been considered? For design, each aspect can influence the choices made in defining a solution. For example: how can the needs of a certain type of user be met? how do the challenges of a particular phase of analysis affect the needs? or, how should the limitations of a display influence the visualization? Similarly, the evaluation of a solution should consider the problem. For example, does it work for the actual users, in the appropriate process phase, and in the context of the real use case? The problem space can also aid as a checklist for documentation: in describing a solution, authors can make clear the problem they were intending to solve by checking they have included at least these aspects.

We believe that a problem space for designing visualizations can provide a useful conceptual tool that can help designers articulate their problem in a manner independent of how it will be solved. We believe this will have benefits in the specification, design (creation of solutions), and evaluation of visualizations.

\section*{ACKNOWLEDGMENTS}
This work has benefitted from Dagstuhl Seminar 22331 "Visualization and Decision Making Design Under Uncertainty”, \url{https://www.dagstuhl.de/22331}.
CG was supported in part by the Federal Ministry of Education and Research of Germany and by the S\"achsische Staatsministerium f\"ur Wissenschaft Kultur und Tourismus in the program Center of Excellence for AI-research "Center for Scalable Data Analytics and Artificial Intelligence Dresden/Leipzig", project identification number: ScaDS.AI.
MG was supported in part by NSF award 2007436. 
MR was supported by the Swedish Research Council – project EXPLAIN VR 2018-03622.
OD was supported in part by Deutsche Forschungsgemeinschaft (DFG, German Research Foundation) – Project-ID 251654672 – TRR 161.
RC was supported in part by NSF awards 1940175,  1939945, and  2118201.

\bibliographystyle{plainnat}
\bibliography{ref-main}


\end{document}

%% file: sidebar.tex
\twocolumn[
\fcolorbox{color1}{white}{
\parbox{\textwidth-1\fboxsep-2\fboxrule}{\centering%
\colorbox{color2!10}{%
\parbox{\textwidth-4\fboxsep-2\fboxrule}{%
\sffamily\textcolor{color1}{\textbf{\\SIDEBAR: TASK ABSTRACTIONS}}\\
\\There is a strong tradition in the field of visualization of developing ways to characterize the needs for visualization in a manner that abstracts the details of particular scenarios. A common approach is to organize scenarios according to \emph{task}, roughly the purpose of the visualization, although the precise definition of task varies widely in the literature \citepSide{rindTaskCubeThreedimensional2016side}.

There are many different types of schemes for organizing tasks, such as taxonomies, typologies, classifications, and categorizations.
While there are technical distinctions between these different terms (see \citetSide{Brehmer2013side}), many researchers use the terms interchangeably, and the subtle distinctions are not relevant to this discussion so we refer to them collectively. 

Early task classification schemes include those by \citetSide{Wehrend1990}, \citetSide{Roth1990}, and \citetSide{Zhou1998}. 
Later, \citetSide{Shneiderman1996} discussed the importance of separating tasks from the type of data.
\citetSide{Amar2005} provided a general set of basic tasks that continues to serve for organizing work on understanding what viewers do with visualizations (e.g., \citetSide{quadriSurveyPerceptionBasedVisualization2021}).
Other key examples include \citetSide{Brehmer2013side}, who introduce the importance of considering tasks at multiple levels, and \citetSide{Schulz2013side}, who introduce the importance of considering multiple facets of tasks. These works already discussed the wide number of available task schemes, which have continued to proliferate.

The works mentioned in the previous paragraph provide general abstractions that apply to all visualizations.
However, many task abstractions focus on specialized subsets. For example, 
\citetSide{leeTaskTaxonomyGraph2006} provide a task taxonomy for graph data, 
\citetSide{kerracherTaskTaxonomyTemporal2015} for temporal graphs, 
\citetSide{Sarikaya2018} for Scatterplots, and 
\citetSide{pandeyStateoftheArtSurveyTasks2022} for trees.
\citetSide{rothEmpiricallyDerivedTaxonomyInteraction2013side} builds a taxonomy for the uses of interaction. 

Two recent works provide important ``meta'' discussions of the literature. \citetSide{Kerracher2017side} describe the processes for constructing task classifications and suggest approaches for evaluating them.  \citetSide{rindTaskCubeThreedimensional2016side} provide an organizational structure for the different task classification schemes, defining a three-dimensional space that different schemes fit into. 

Our work builds on this existing literature on task abstraction by recognizing it as one of many aspects that must be considered in designing a visualization. 
\small{
\bibliographystyleSide{plainnat}
\bibliographySide{ref-sidebar}}
}}}}]